\documentclass[onecolumn,superscriptaddress]{revtex4} 
\usepackage{graphicx}
\usepackage{amsmath,amssymb}

\begin{document}
\title{Nanoscale confinement of ultrafast spin transfer torque exciting non-uniform spin dynamics by femtosecond spin current pulses}


\author{Ilya Razdolski}
\email{razdolski@fhi-berlin.mpg.de}
\affiliation{Physical Chemistry Dept., Fritz Haber Institute of Max Planck Society, Faradayweg 4-6, 14195 Berlin, Germany}
\author{Alexandr Alekhin}
\affiliation{Physical Chemistry Dept., Fritz Haber Institute of Max Planck Society, Faradayweg 4-6, 14195 Berlin, Germany}
\author{Nikita Ilin}
\affiliation{Physical Chemistry Dept., Fritz Haber Institute of Max Planck Society, Faradayweg 4-6, 14195 Berlin, Germany}
\author{Jan~P.~Meyburg}
\affiliation{Faculty of Chemistry, University of Duisburg-Essen, Universit\"{a}tsstr. 5, 45141 Essen, Germany}
\author{Vladimir~Roddatis}
\affiliation{Universit\"{a}t G\"{o}ttingen, Institut f\"{u}r Materialphysik, Friedrich-Hund-Platz 1, 37077 G\"{o}ttingen, Germany}
\author{Detlef~Diesing}
\affiliation{Faculty of Chemistry, University of Duisburg-Essen, Universit\"{a}tsstr. 5, 45141 Essen, Germany}
\author{Uwe Bovensiepen}
\affiliation{Faculty of Physics and Center for Nanointegration (CENIDE), University of Duisburg-Essen, Lotharstr. 1, 47057 Duisburg, Germany}
\author{Alexey Melnikov}
\email{melnikov@fhi-berlin.mpg.de}
\affiliation{Physical Chemistry Dept., Fritz Haber Institute of Max Planck Society, Faradayweg 4-6, 14195 Berlin, Germany}

\date{\today}

\maketitle \textbf{
Spintronics had a widespread impact over the past decades due to transferring information by spin rather than electric currents \cite{Slonczewski96,Berger96,Kiselev03,Krivorotov05,Woltersdorf07,Brataas12}. Its further development requires miniaturization and reduction of characteristic timescales of spin dynamics combining the sub-nanometer spatial and femtosecond temporal ranges \cite{BaderParkin}. 
These demands shift the focus of interest towards the fundamental open question of the interaction of femtosecond spin current (SC) pulses with a ferromagnet (FM). 
The spatio-temporal properties of the impulsive spin transfer torque (STT) exerted by ultrashort SC pulses on the FM open the time domain for probing non-uniform magnetization dynamics.
Here we employ laser-generated ultrashort SC pulses \cite{Malinowski08,Melnikov11,SpinSeebeck} for driving ultrafast spin dynamics in FM and analyzing its transient local source.
Transverse spins injected into FM excite inhomogeneous high-frequency spin dynamics up to 0.6 THz, indicating that the perturbation of the FM magnetization is confined to 2 nm.
%
}

\begin{figure}[b]
	\includegraphics[width=0.5\columnwidth]{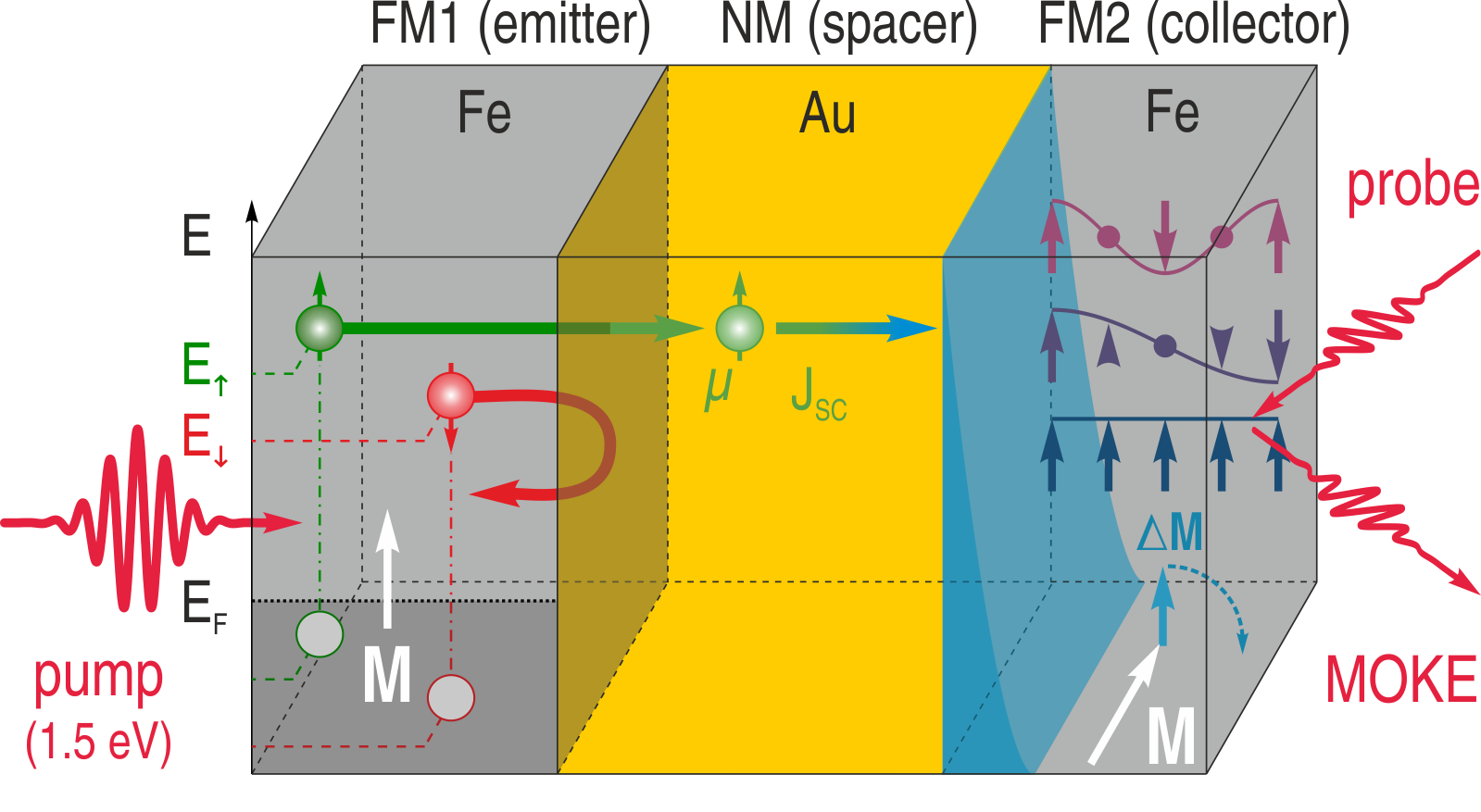}
	\caption{ {\bf Laser-induced excitation of spin dynamics via spin current pulses.}
	Laser pump pulse impinging on the first Fe layer (emitter) excites hot electrons at elevated energies $E_{\uparrow,\downarrow}$.
	Because of the unequal energies and therefore different transmittance of the Fe/Au interface for the majority (green) and minority (red) hot electrons \cite{SpinSeebeck},
	the emission of hot electrons into Au is largely spin-polarised. 
	Having crossed the Au layer in a nearly ballistic regime, the electrons reach the FM2 layer and transfer their spin to it.
	Owing to that, STT is exerted on the magnetization ${\bf M}_{\rm C}$ which is driven out of the equilibrium and starts precessional dynamics. 
Due to the spatial confinement of the STT perturbation \cite{StilesPRB02} (blue shaded area), spin waves with a broad spectrum of non-zero wavevectors $k$ are excited and can be probed by the magneto-optical Kerr effect in the collector.
	}
\end{figure}

Approaching the timescales of the underlying elementary processes, SCs with femtosecond pulse duration can provide valuable fundamental insights into the ultrafast spin dynamics. In addition to manipulating the magnetization in multilayer structures \cite{Rudolf12,Choi14,Choi15}, ultrashort SC pulses were shown to exert STT and thus drive the coherent magnetization dynamics in semiconductor films \cite{Nemec12} or perpendicularly coupled magnetic bilayers \cite{Schellekens14,Choi15}. 
However, a nearly uniform in-depth STT limited the coherent magnetization dynamics to the homogeneous precession (with $k=0$) on the picosecond timescale, similarly to other known ultrafast excitation mechanisms
\cite{Ju99,Zhang02,Kimel05,Hansteen05,Afanasiev14}.
An inhomogeneous perturbation of magnetization mediated by the interfacial STT contribution \cite{SpinSeebeck,StilesPRB02} can excite spin waves in a FM film, thus
extending the spin dynamics into higher frequency range.
Moreover, the spatial properties of the SC-driven STT excitation can be inferred from the spectral analysis of these high-frequency spin waves. 
In this Letter, we realize this approach in epitaxial Fe/Au/Fe/MgO(001) multilayers by means of all-optical excitation and detection of the standing spin waves in a 15-nm thick Fe film via the STT mechanism (see Fig.~1). Further, we demonstrate the excited complex mode structure of the non-uniform magnetization dynamics. We show that the ultrashort laser-induced SC pulses constitute a convenient tool to excite spin waves and study the interaction of spins with a non-collinear magnetization.

The concept of our experiment is illustrated in Fig.~1. A 14 fs long laser pump pulse absorbed in the FM1 layer (emitter, thickness 16 nm) results in the emission of the subpicosecond SC pulse into Au via the non-thermal spin-dependent Seebeck effect \cite{SpinSeebeck}. Owing to their large lifetimes in Au
\cite{Zhukov06},
the transport of hot electrons in a quasi-ballistic regime delivers spin angular momentum to the FM2 layer (collector, 14~nm-thick).
With this spin polarization orthogonal to the collector magnetization ${\bf M}_{\rm C}$, both reflected and transmitted electrons in the SC pulse transfer the transient angular momentum density ${\bf \mu }(t)$ to FM2 and thus exert a STT on ${\bf M}_{\rm C}$ \cite{StilesPRB02}. Locally, the interaction of ${\bf \mu }(t)$ with the magnetization ${\bf M}_{\rm C}$ is given by: \cite{Slonczewski96}

\begin{equation}\label{LLtorque}
	\frac{1}{\gamma}\frac{\partial {\bf M}_{\rm C}}{\partial t}\bigg\rvert_{\rm STT}= \lambda{\bf M}_{\rm C}\times[\vec{\mu}(t)\times{\bf M}_{\rm C}].
\end{equation}

\begin{figure}[t]
	\includegraphics[width=0.5\columnwidth]{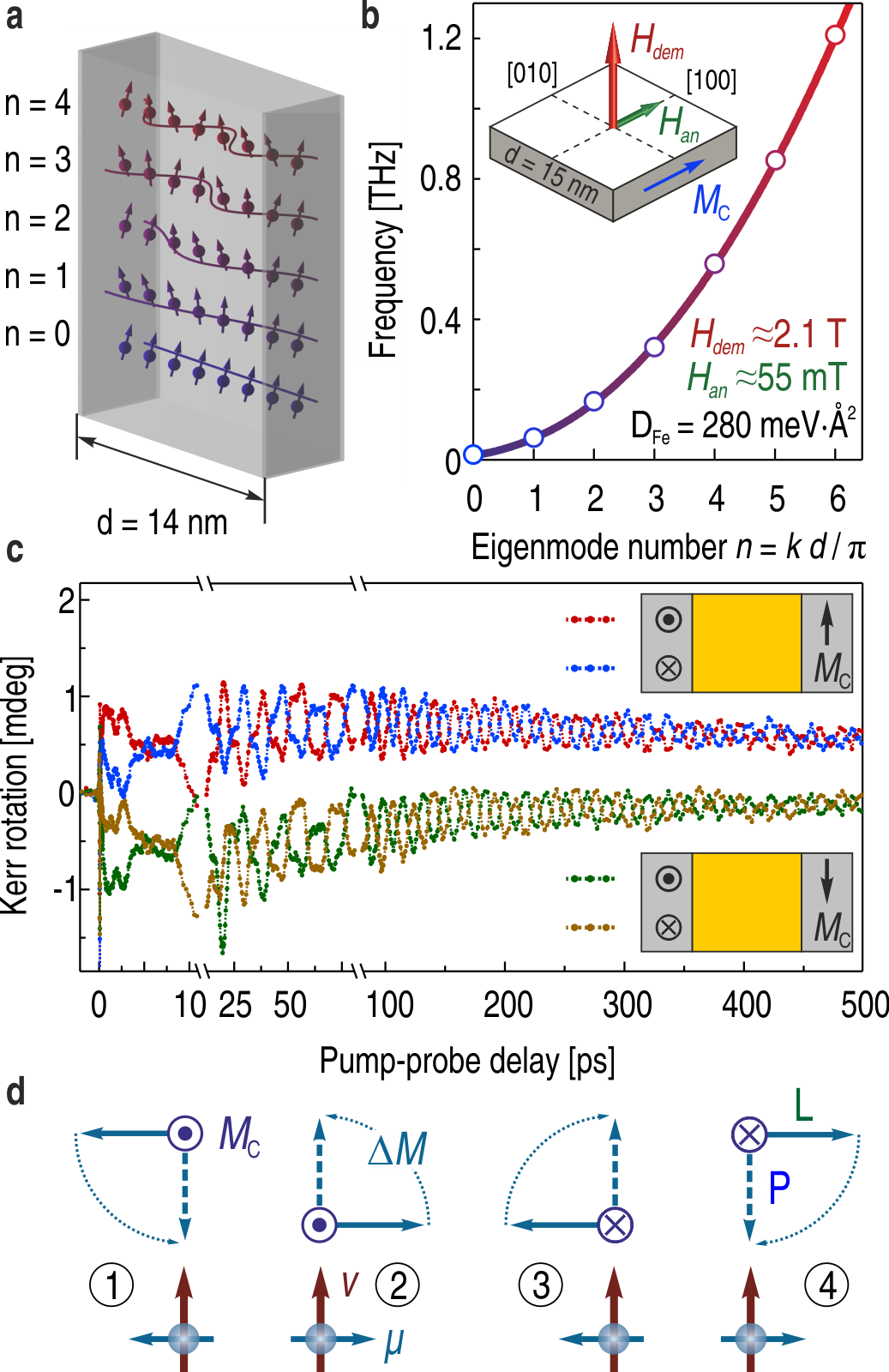} 
	\caption{{\bf Laser-induced magnetization dynamics in the Fe/Au/Fe/MgO(001) multilayer.} 
	{\bf a, b}, Spin excitations in a thin Fe(001) film magnetized in-plane.
	The demagnetizing field $H_{\rm dem}$ and the two in-plane easy axes with the effective crystalline anisotropy field $H_{\rm an}$ result in the uniform magnetization precession with a frequency $f_0\approx 10$~GHz \cite{Carpene10}. A film with the thickness $d$ and the exchange stiffness $D_{\rm Fe}$ can sustain standing spin waves with a discrete set of eigenwavevectors $k_n=\pi n/d$ and frequencies $f_n$ given by Eq.~(2).
	{\bf c}, Time-resolved transient polar MOKE rotation obtained in four different geometries for orthogonal magnetizations of the emitter and collector. The periodic signals with multiple frequencies indicate the excitation of the standing spin-wave eigenmodes.
	{\bf d}, Separation of the longitudinal and polar MOKE contributions. Electrons in the SC pulse propagate with the velocity $v$ and magnetic moment $\vec{\mu}$. For the longitudinal SC polarization $\vec{\mu}$ and transverse magnetization of the collector ${\bf M}_{\rm C}$, the solid (dashed) blue arrows depict the sign of the longitudinal, L (polar, P) transient magnetization component $\Delta M$. The dotted lines illustrate the appearance of the polar component of ${\bf M}_{\rm C}$ due to the precession of magnetization. 
	}
\end{figure}

The ultrafast impulsive STT excitation triggers spin dynamics in the collector. Because of the strong localization of the delivered perturbation in the vicinity of the Au/Fe interface \cite{SpinSeebeck,StilesPRB02}, spin waves with non-zero $k$-vectors are excited along with the homogeneous precession of magnetization ${\bf M}_{\rm C}$ (Fig.~2,a). 
The spin-wave dispersion for a thin magnetic film (Fig.~2,b) is given by \cite{GurevichMelkov}:

\begin{equation}\label{dispersion}
f(k)=\gamma\sqrt{(H_{\rm an}+Dk^2)\cdot(H_{\rm an}+H_{\rm dem}+Dk^2)},
\end{equation}
where $H_{\rm dem}
\approx~ 2.1$~T
(Fig.~2,b), $\gamma\approx~28$~GHz/T is the gyromagnetic ratio, and $D_{\rm Fe}=280$~A$^2\cdot$~meV
\cite{Mook73}.
In a film of a finite thickness, only the standing waves with $k_n=\pi n/d$  are supported (Fig.~2,a-b), where the zero derivative of the magnetic moment at the interfaces \cite{GurevichMelkov} is ensured by the low Fe/Au interface anisotropy.
Rich dynamics of the time-resolved MOKE signals observed in our experiments (Fig.~2,c) is associated with the superposition of these long-lived standing spin-wave modes.

\begin{figure*}[t]
	\includegraphics[width=0.95\textwidth]{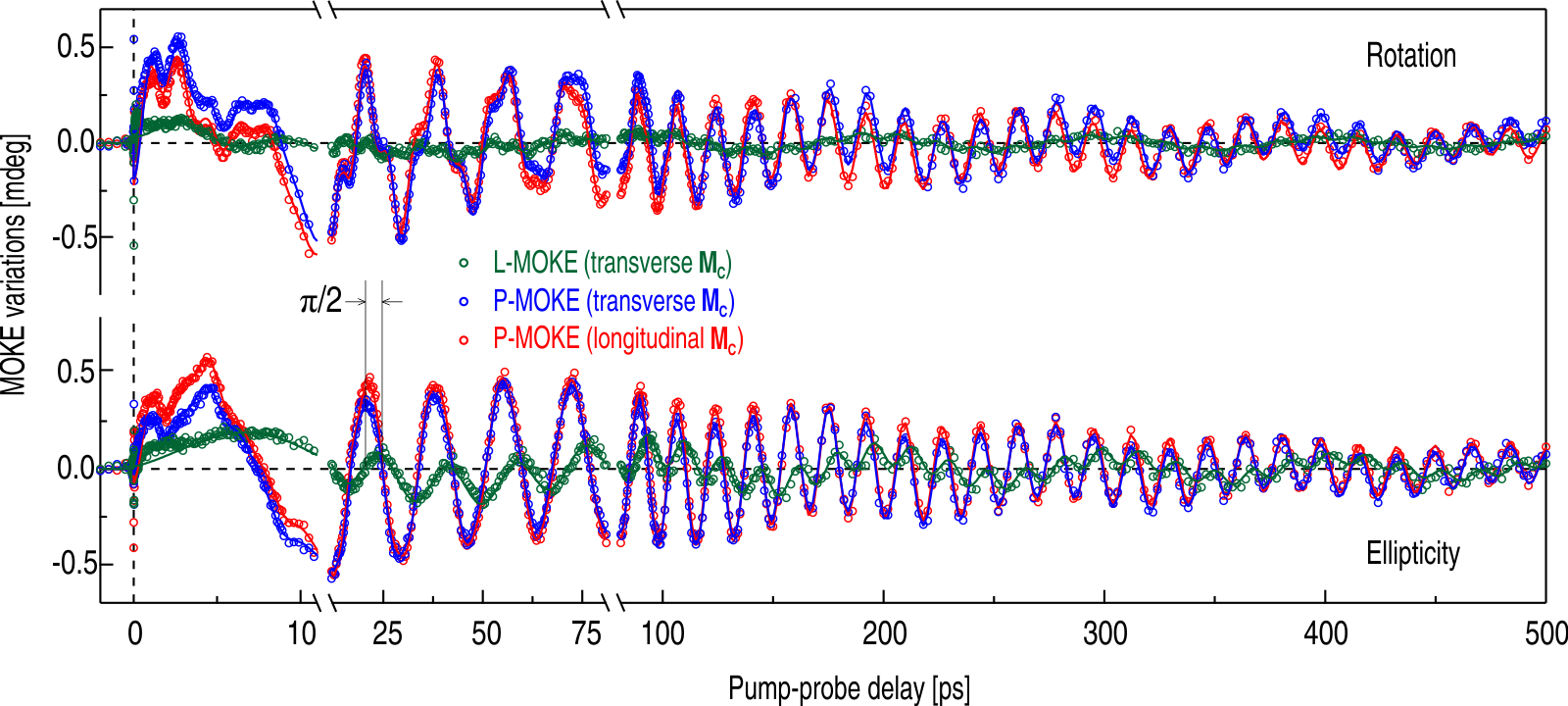} 
	\caption{ {\bf STT-induced magnetization dynamics.} 
	Transient MOKE signals in both rotation (top) and ellipticity (bottom). Polar and longitudinal MOKE effects are separated. Solid lines are the results of the fitting procedure with a set of exponentially decaying oscillations with their frequencies given by Eq.~(2).
The pronounced similarity between the two polar MOKE dynamics (red and blue dots) demonstrates the robustness of the measurements. The indicated $\pi/2$~phase shift between the polar and longitudinal MOKE components confirms the precessional nature of the magnetization dynamics.
		}
\end{figure*}

In lieu of fitting the raw data, we performed measurements in various combinations of the magnetization directions of FM1 and FM2, while keeping them perpendicular to each other. Based on the parity rules with respect to both $\vec{\mu}$ and ${\bf M}_{\rm C}$ (see Supplementary Note 1), we disentangle the polar and longitudinal MOKE contributions to magnetization dynamics in the FM2 layer (Fig.~2d). 
This procedure results in background-free data for MOKE rotation and ellipticity (Fig.~3).

It is clearly seen that the transient MOKE signals have a complex structure demonstrating oscillations at multiple frequencies. To unravel their nature, we performed the Fourier analysis of the time-resolved MOKE data. The analysis shows the presence of four frequencies (Fig.~4,a) besides the fundamental excitation which corresponds to the uniform magnetization precession ($k=0$). We argue that these frequencies indicate the excitation of the long-lived (up to 500 ps) standing spin waves in the FM2 film. 
The striking match between the frequencies obtained from the Fourier analysis and those calculated from the standing spin-wave dispersion illustrated in Fig.~4,a verifies our explanation.
We fitted a set of the exponentially decaying oscillations with the five frequencies given by Eq.~(\ref{dispersion}) to the experimental data. The excellent quality of the fitting results corroborate our understanding of the standing spin-wave excitation (Fig.~3, solid lines). The difference between the MOKE rotation and ellipticity data, as well as between the polar and longitudinal MOKE effects is attributed to their unequal sensitivity to the in-depth magnetization profile $m(z)$~\cite{HamrleJPD10,WieczorekPRB15} and thus to various standing spin-wave eigenmodes (see Supplementary Note 2).

These data unambiguously prove that the STT mechanism is capable of exciting high-frequency modes of spin precession in FM films.
Note that because the data shown in Figs.~2,3 were obtained in the absence of an external magnetic field, the heating mechanism of the excitation of the magnetization precession relying on the ultrafast quenching of the magnetic anisotropy is inactive (see Supplementary Note 3).

In order to elucidate the STT-induced excitation, we now turn to the ultrafast timescale at $t<1$ ps. Figure~4,b shows the initial stage of the STT-induced magnetization dynamics, when the emitter and collector films are magnetized longitudinally and transversely, respectively.
Due to angular momentum conservation, the longitudinal spin polarization of the SC pulse drives the rapid initial increase of the corresponding magnetization projection starting at approximately 50~fs delay, indicating the ballistic SC propagation through the 55~nm-thick Au spacer \cite{Melnikov11}.
We found the duration of the SC pulse $\tau_{\rm SC}\lesssim 250$ fs, in agreement with the results of the direct SC measurements \cite{SpinSeebeck}. 
On the picosecond timescale only resonant spin waves from the initially excited wavepacket with a broad distribution of $k$-vectors endure. Thus, standing spin waves in the collector are formed, giving rise to the oscillatory dynamics seen in Fig.~4,b.

\begin{figure*}[t]
	\includegraphics[width=0.95\textwidth]{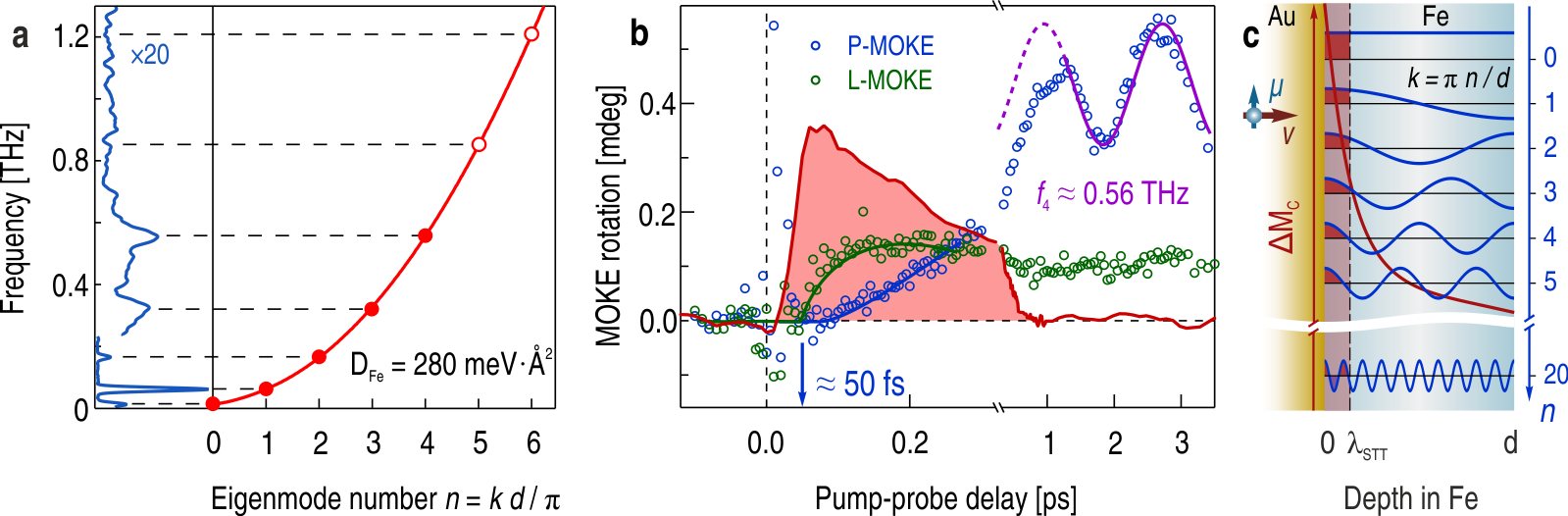} 
	\caption{ {\bf Excitation of the standing spin waves.}
	{\bf a}, Fourier spectrum of the experimental data, averaged over several datasets (left panel) and the frequencies of the standing spin waves in a 15 nm-thick Fe film (right panel, red symbols). Red solid line is the calculated spin waves dispersion curve from Eq.~(\ref{dispersion}) with the indicated magnon stiffness $D_{\rm Fe}$ . Along with the fundamental uniform precession ($n=0$), four higher modes are shown which were detected in the experiment (full dots).
	{\bf b}, Transient MOKE rotation data on the ultrashort timescale. The rapid onset of the longitudinal component (green circles) is followed by a slower increase of the polar one (blue circles) due to the magnetization precession around its equilibrium. The red shaded area reproduces the SC pulse as measured in Ref.~\cite{SpinSeebeck}. The purple line illustrates the 4$-th$ spin wave eigenmode with the frequency $f_4$ observed in the transient polar MOKE component.
	{\bf c}, Excitation of the higher eigenmodes is limited by the spatial in-depth profile of the STT perturbation ${\bf M}_{\rm C}$ (red curve). 
	The excitation region with a characteristic depth $\lambda_{\rm STT}$ is sketched by the red shaded area.
	The excitation efficiency given by the convolution of the STT excitation profile and the in-depth profile of a particular standing spinwave eigenmode (blue curves) is represented with red and blue shaded areas. In the case of a homogeneous precession (the top eigenmode with $n=0$), this overlap is always non-zero. The eigenmodes with small non-zero $n$ are still excited albeit with smaller efficiency. However, for the eigenmode with a large $n$ (bottom, $n=20$), the convolution yields positive and negative regions (red and blue shaded areas) which cancel each other out, thus greatly reducing the excitation efficiency. 
		}
\end{figure*}

The transverse angular momentum is transferred from the injected electrons to the magnetization ${\bf M}_{\rm C}$ in the vicinity of the interface, locally driving magnetization dynamics according to Eq.~1. Thus, besides the SC pulse duration $\tau_{\rm SC}\lesssim 250$ fs allowing the excitation of the modes with frequencies up to $1/2\tau_{\rm SC}\approx 2$ THz, the spectrum of the excited magnons is limited by the $k$-vector spectrum of the delivered excitation, or, in other words, the STT characteristic depth $\lambda_{\rm STT}$.
In Fig.~4,b it is seen that the first detected oscillation is the $n=4$ mode with a frequency of $f_4=0.56$~THz. As such, the $n=5$ and the higher modes are supposedly not excited, due to either spatial or temporal limitations.
The estimated $\tau_{\rm SC}$ complies with the requirement for the efficient impulsive excitation of the $n=5$ eigenmode, $\tau_{\rm SC}<T_5/2\approx 0.6$~ps.
Thus, the temporal constraint can be excluded and we need to invoke the spatial inhomogeneity-driven limitation on the excited eigenmodes.
Figure~4,c illustrates that for the standing spin waves with open ends, the critical STT excitation depth is about a quarter of the wavelength. This means that the eigenmode with $k_5$ is not efficiently excited if 
$\lambda_{\rm STT}> 1/4\times 2\pi/k_5\approx 1.5$~nm.

We note that there are other possible limitation mechanisms related to the lifetime of the eigenmodes. From the fitting procedure, the eigenmode with $k_4=4\pi/d$ was found to live for about 6~ps. The eigenmodes with even larger $k$ most likely have even shorter lifetimes (comparable to their periods or shorter) and thus render invisible in our experiments.
However, the very fact that the eigenmode with $k_4$ is unambiguously observed in the experiments can be used for a rather conservative estimate
$\lambda_{\rm STT}\lesssim 1/4\times 2\pi/k_4\approx2$ nm. 
Together with the L-MOKE data shown in Fig.~4b, this value indicates that the angular momentum transfer to the FM2 layer results in a 1.3$^{\circ}$ tilt of ${\bf M}_{\rm C}$ from its equilibrium direction within $\lambda_{\rm STT}$.

With the ultimately short (2~nm) characteristic depth, the hot electron-induced STT remains one the most efficient mechanisms for the excitation of the non-uniform magnetization dynamics with large $k$-vectors. For comparison, the optical penetration depth in transition metals is of the order of the skin depth $\delta\approx 10-15$ nm. In this case, the excitation of the standing spin waves with non-zero $k$-vectors would become possible in relatively thick films only, which effectively reduces the eigenmodes frequencies down to the tens of GHz \cite{Carpene10,vanKampen02}. 
In this regard, the ultrashort laser-induced pulses of SC
are a unique tool capable of exciting large-$k$, sub-THz spin waves in FM films.
Our results demonstrate the extreme abilities of the SC pulses at exciting non-uniform spin dynamics and elucidate the interaction of the SC with a non-collinear magnetization. We found that on average, a magnetic moment of about 1$\mu_B$ per Fe atom at the Au/Fe interface is transferred to the collector, showing high promise for the magnetization switching in thin FM layers.
With these results in hand, further steps towards ultrafast spintronics can be expected in the near future.

{\bf Methods}\\
{\small
{\bf Sample fabrication.} Epitaxial Fe/Au/Fe structures with 15~and~16~nm Fe layers and 55~nm Au spacer capped with
3 nm of Au were grown on optically transparent MgO(001) substrates with a thickness of 0.5 mm. The substrates were cleaned in the ultrasonics baths of ethanol, isopropanol, acetone (15~min each at elevated temperature of $\approx$ 310~K). After being mounted in a UHV chamber, the substrates were exposed to oxygen with a partial pressure of $2\times10^{-3}$~mbar at a temperature of 540~K for 30~min in order to remove spurious amounts of diamond polish left by the last fabrication step. With UHV conditions reached, Fe and the first nm of the interstitial Au layer were evaporated at 460 K. The samples were then cooled down and the additional amount of Au was evaporated at room temperature.
Transparent substrate and thin capping Au layer provided optical access from both sides. The scanning transmission electron microscopy revealed excellent epitaxial quality and flatness of interfaces \cite{SpinSeebeck} (see Supplementary Note 4).}\\

{\small
{\bf Magnetic characterization.} 
Prior to the time-resolved experiments, magnetic properties of the samples were characterized by means of the static MOKE measurements. The hysteresis loops of the MOKE rotation and ellipticity were obtained from both sides of the samples (see Supplementary Note 5). The sample was placed in such a way that its easy anisotropy axes were oriented along the longitudinal and transverse magnetic field directions. During the measurements, a main magnetic field was scanned (from -20~to~20~Oe) across the longitudinal direction, while a weak auxiliary magnetic field (up to 5~Oe) was applied in the transverse direction.  Both fields were produced by electromagnets. This procedure allowed us to realize the two-step switching of the magnetization from one longitudinal direction to another via the intermediate transversal direction (along the auxiliary magnetic field). In the MOKE measurements on the collector and emitter films, we obtained different width of the hysteresis loops. This allowed us to attain an orthogonal magnetic state, where the magnetizations of the collector and the emitter were perpendicular to each other.
}

{\small
{\bf Time-resolved measurements.} In our back pump-front probe experiments we used p-polarized 800~nm, 14~fs output of a commercial cavity-dumped Ti:sapphire oscillator (Mantis, Coherent) operated at 1 MHz, which was split at a power ratio 4:1 into
pump and probe pulses. Both beams were chopped with different frequencies and focused with off-axis parabolic
mirrors to about 10~$\mu$m spot size, resulting in a pump fluence of the order of 10~mJ/cm$^2$. The signal reflected from the sample was split into two identical shoulders in order to allow for simultaneous measurements of the MOKE rotation and ellipticity. Balanced detection scheme in each shoulder was realized with the help of a Glan-laser prism and two photodiodes. In the ellipticity shoulder a quarter-wave plate was installed before the prism. The measurements were performed at room temperature. The zero time delay was determined before each measurement with the help of the cross-correlation second harmonic generation signal.
}

\textbf{Acknowledgements}
The authors are indebted to A. Kirilyuk, S. Sanvito, S. O. Demokritov, M. Weinelt, P. Oppeneer and T. Kampfrath for fruitful discussions, as well as M. Wolf for the financial support. This work was partially funded by the Deutsche Forschungsgemeinschaft (ME 3570/1, Sfb 616) and by the EU 7-th framework program (CRONOS).
\\



\textbf{Correspondence}\\Correspondence and request for materials should be addressed to I.R. or A.M.\\


\newpage

\begin{center}
\huge{Supplementary Information}
\end{center}

\section*{Supplementary Note 1: Separation of the polar and longitudinal MOKE contributions}

The spin transfer torque (STT)-induced contribution to the magnetization dynamics is given by \cite{SlonczewskiJMMM96}:

\begin{equation}\label{LLtorque}
	\frac{1}{\gamma}\frac{\partial {\bf M}}{\partial t}\bigg\rvert_{\rm STT}= \lambda{\bf M}\times[\vec{\mu}(t)\times{\bf M}].
\end{equation}

Here $\lambda$ is the scaling factor, $\vec{\mu}$ is the magnetic moment carried by the spin current and ${\bf M}$ is the magnetization of a ferromagnet. It is seen that the STT-induced magnetization dynamics has different parity with respect to both $\vec{\mu}$ and ${\bf M}$. Whereas variations of ${\bf M}$ are odd with respect to $\vec{\mu}$, at the same time they are even with respect to ${\bf M}$ itself. In other words, consider the collector magnetized transversely along the $y$-axis ($M_y\neq 0$), and the emitter magnetization (proportional to $\vec{\mu}$) along the $x$-axis (longitudinal). Here, upon the impulsive STT excitation the collector magnetization ${\bf M}$ acquires a $x$-projection $M_x$. The sign of $M_x$ replicates the sign of $\mu_x$ and is independent of the sign of $M_y$. 

Further, the magnetization precession around its equilibrium (parallel to the $y$-axis) leads to the emergence of the dynamic polar (out-of-plane) component $M_z$. Moreover, at the initial stage of the precession $\Delta M_z\propto dM_z/dt \propto H_{\rm an}M_x$, where $H_{\rm an}$ is the anisotropy field parallel to the $y$-axis and thus the non-perturbed direction of the magnetization $M_y$. It is seen that the sign of the polar component $M_z$ is determined by both signs of the projections $\mu_x$, $M_y$, whereas the sign of the longitudinal component $M_x$ depends on the sign of the $\mu_x$ only. Owing to that, it becomes possible to separate the polar MOKE contribution from the longitudinal one when dealing with four datasets measured in various configurations of the equilibrium magnetizations, $\pm\mu_x$, $\pm M_y$ (Fig.~2c-d in the main Manuscript). Indeed, consider the notation where U, D (R, L) as in up, down (right, left) denote the magnetization and magnetic moment directions parallel and antiparallel to the $y$ ($x$) axis. The experimental configurations with two orthogonal magnetizations discussed above can be written as RU, RD, LU, LD, where the first and the second symbols represent the magnetization of the emitter (proportional to $\mu$) and the collector $M$, respectively.
As such, odd with respect to $\mu$, the pure longitudinal dynamics of Kerr rotation is given by 

$$
\varphi_{L}=\frac{1}{4}(\varphi_{\rm RU}+\varphi_{\rm RD}-\varphi_{\rm LU}-\varphi_{\rm LD}),
$$
Further, odd (even) with respect to the magnetization of the emitter (collector), the polar Kerr dynamics is given by
$$
\varphi_{P}=\frac{1}{4}(\varphi_{\rm RU}-\varphi_{\rm RD}+\varphi_{\rm LU}-\varphi_{\rm LD}).
$$  

%
%

\section*{Supplementary Note 2: MOKE in-depth sensitivity}

\begin{figure}[h]
		\includegraphics[width=0.7\textwidth]{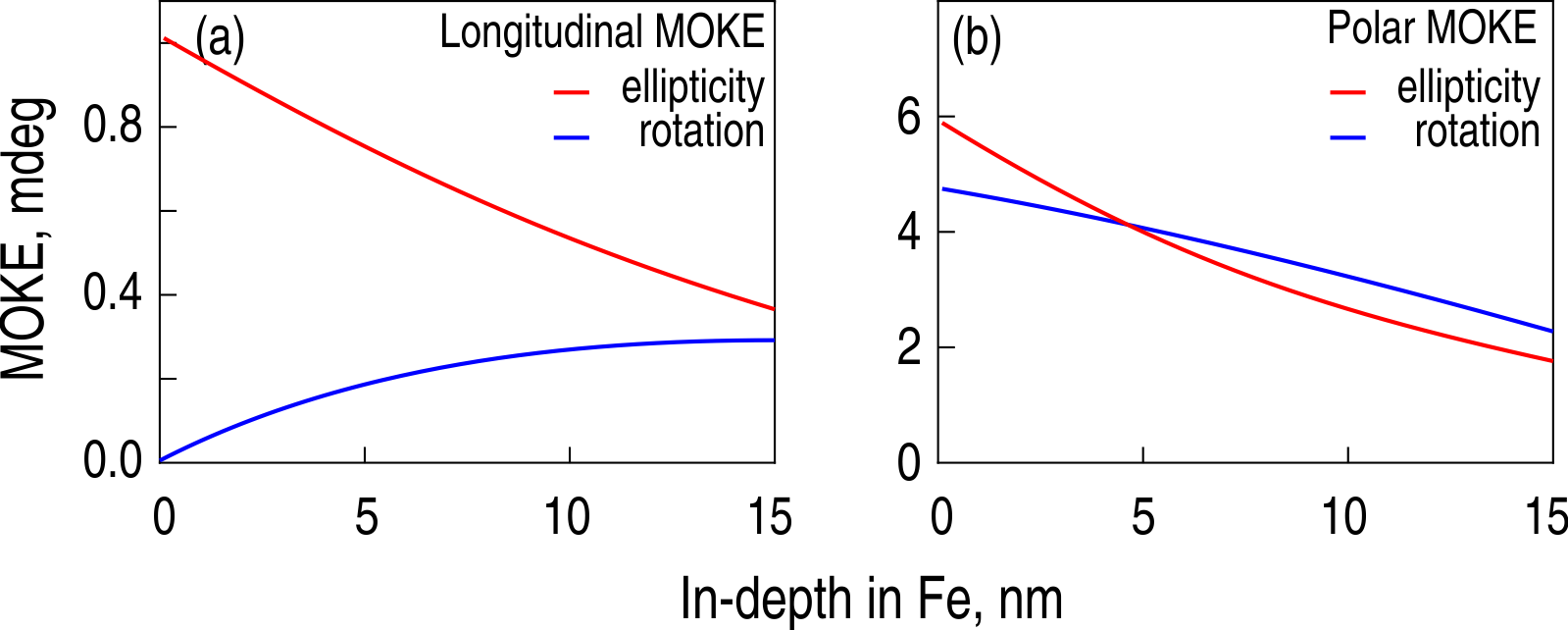}
			\caption{ 
	Sensitivity $s(z)$ of (a) longitudinal and (b) polar MOKE rotation (full blue circles) and ellipticity (open red circles) in a 15 nm-thick Fe film for the p-polarized light. 
}
\label{sens}
\end{figure}

Despite having no total magnetic moment, standing spin waves can be visualized in MOKE experiments because of the in-depth selective sensitivity of MOKE. Figure~\ref{sens} illustrates the calculated specific MOKE rotation and ellipticity for the 15 nm-thick Fe films. The data were calculated using the layer-by-layer approach based on the medium boundary and propagation matrices \cite{Zak90,Traeger92,HamrleNJP10,WieczorekPRB15} for the p-polarized incident beam. 

The total MOKE response $\theta,\varepsilon$ is given by a convolution of the sensitivity $s(z)$ and the certain in-depth profile of magnetization $m(z)$:

\begin{equation}
\label{sensitive}
\theta,\varepsilon = \int_0^d{s_{\rm \theta,\varepsilon}(z)m(z)dz}
\end{equation}

Two main considerations can be inferred from the data shown in Fig.~\ref{sens}. Firstly, it is seen that longitudinal MOKE rotation and ellipticity are predominantly sensitive to the opposite interfaces of the film. On the contrary, polar MOKE rotation and ellipticity exhibit very similar sensitivities to local magnetization $m(z)$. As such, the differences in the transient rotation and ellipticity in the case of the longitudinal MOKE could be attributed to a different dynamics at the two interfaces of the Fe film.


\section*{Supplementary Note 3: Separation of the heating and STT contributions in the spin waves excitation}

The experiments described in the main Manuscript were performed without external magnetic field. In order to study the heating effect on the transport-induced magnetization dynamics in Fe/Au/Fe trilayers, we performed additional measurements with an external magnetic field $\vec{H}$ on. This magnetic field was applied in the plane of the sample at an angle with respect to the easy axes of the two Fe films, so that the magnetizations of the emitter and the collector were aligned orthogonally to each other (see Supplementary Note 5).

\begin{figure}[h]
		\includegraphics[width=0.95\textwidth]{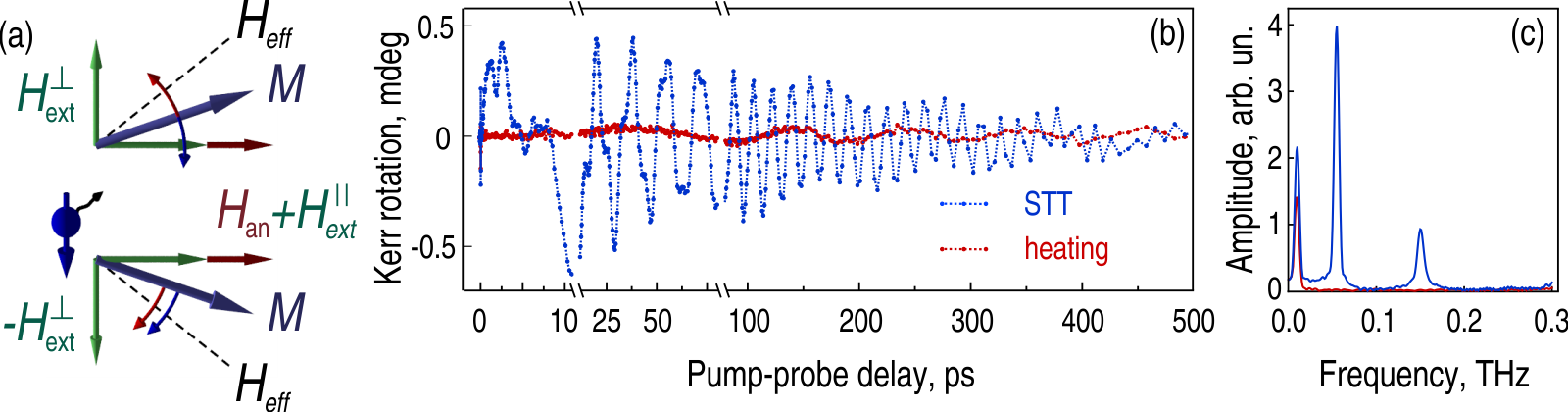}
	\caption{ 
	(a) Schematic of the heating and STT-induced magnetization dynamics for two opposite directions of $H_{\rm ext}^{\perp}$, a projection of the external magnetic field perpendicular to the magnetization $M$ of the collector. Red and blue short arrows depict the initial dynamics of the magnetization induced by the heating and STT mechanisms, respectively.
	(b-c) Heating- (red) and STT (blue)-induced contributions to the polar time-resolved MOKE signal (b) and their spectra (c).
}
\label{heating}
\end{figure}

In this configuration (Fig.~\ref{heating},a), the equilibrium direction of the collector magnetization $M$ is determined by an interplay of the external magnetic field $\vec{H}_{\rm ext}=\vec{H}_{\rm ext}^{\|}+\vec{H}_{\rm ext}^{\perp}$ and the anisotropy field $\vec{H}_{\rm an}$. The thermal action of the ultrashort spin current pulse results in heating of the collector and thus quenching its magnetic anisotropy $H_{\rm an}$. As such, the equilibrium direction of the magnetization $M$ shifts giving rise to the precessional dynamics around $H_{\rm eff}$ with the frequency $f_0\approx 10$ GHz \cite{Carpene10}. 

Figure~\ref{heating},a illustrates this mechanism along with the STT excitation discussed above. Note that according to the Eq.~(\ref{LLtorque}), the STT (heating) mechanism is even (odd) with respect to the external field ${\bf H_{\rm ext}}$. 
Thus, the direction of the heating-induced precession reverses together with the perpendicular projection applied external field ${\bf H_{\rm ext}^{\perp}}$. We note, however, that the STT mechanism does not depend on the external magnetic field as long as the directions of the magnetizations of both emitter and collector are set. Thus, obtaining the MOKE data for the two opposite projections of the external field ${\bf H_{\rm ext}^{\perp}}$ perpendicular to the equilibrium collector magnetization (but keeping the magnetization $M$ in place), we can separate the heating and STT contributions (Fig.~\ref{heating},b). Whereas the STT-driven dynamics (blue symbols) has a much larger amplitude and can boast a rich spectrum of frequencies, the heating-induced magnetization dynamics (red symbols) is small and consists of a single FMR mode only. The latter can be explained invoking the good heat conductivity of Fe along with the large duration of heat pulses as compared to spin current ones. Both these factors facilitate the excitation of the homogeneous precession with $k=0$ (FMR) and inhibit any inhomogeneous (with $k>0$) dynamics of thermal origin. 

\section*{Supplementary Note 4: Samples transmission electron microscopy}

\begin{figure}[!h]
		\includegraphics[width=0.7\textwidth]{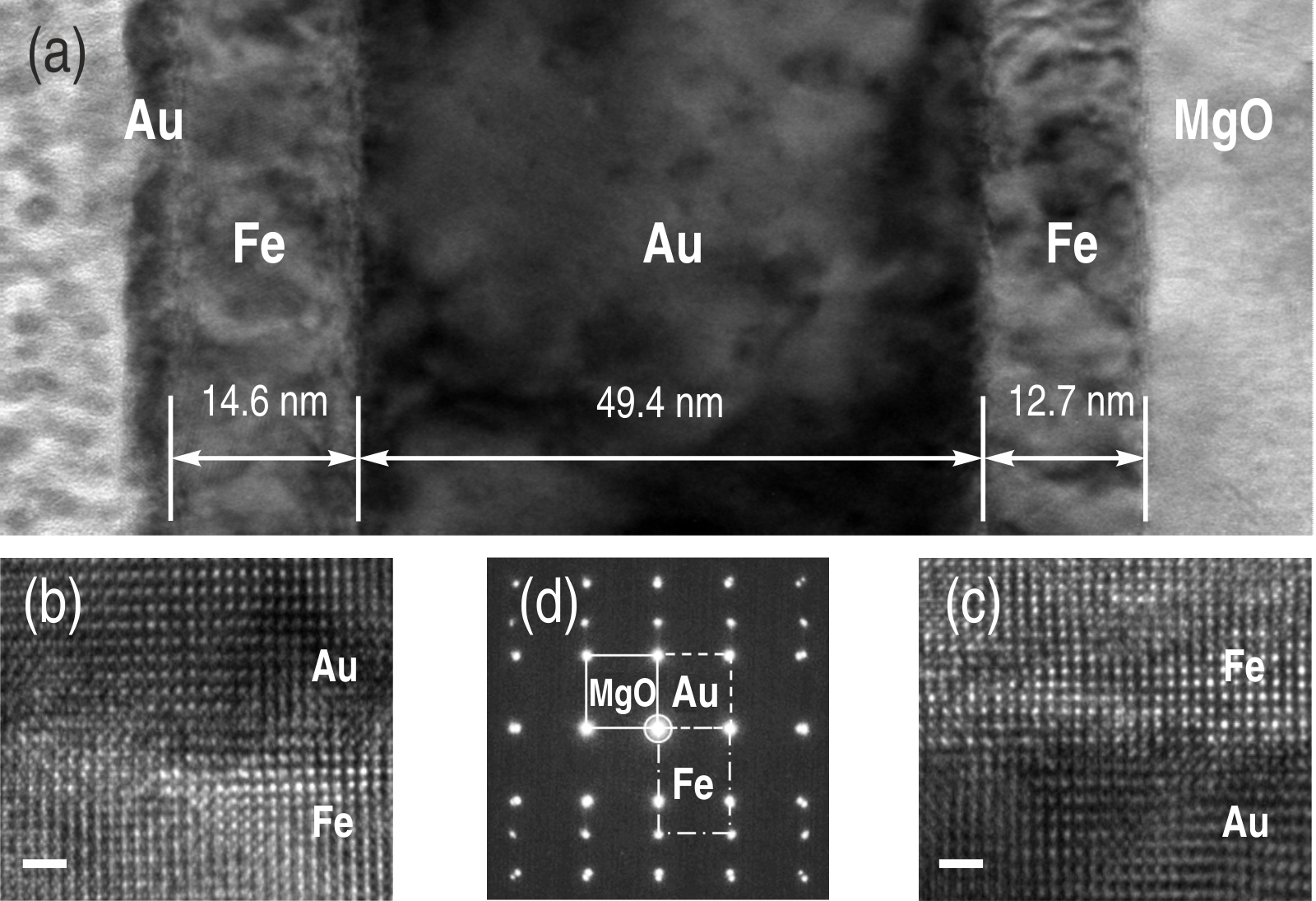}
			\caption{ 
	(a) Au/Fe/Au/Fe/MgO(001) cross-section TEM image.
	(b-c) High-resolution TEM images of flat Fe/Au and Au/Fe interfaces. The horizontal bar is 1 nm long.
	(d) Electron diffraction image.
}
\label{microscopy}
\end{figure}

The Fe/Au/Fe/MgO(001) samples were capped with a 3 nm-thick Au protective layer. The structure of the samples was characterized using a transmission electron microscope (TEM) TITAN 80-300 (FEI, USA) equipped with a corrector of spherical aberration at image side. The microscope was operated at $300$~kV. An example of cross-section specimen prepared by means of the Focused Ion Beam lift-out technique is shown in Fig.~\ref{microscopy}. The cross-section was prepared at the spot used for the optical measurements. The micro-structural study reveals that both Fe and Au films grow epitaxially and the roughness of Fe/MgO and Fe/Au interface varies from 0.7 to 1.5~nm. The precise measurements of the layers thicknesses yield the values of $12.7$ and $14.6$~nm for the Fe collector and emitter, respectively, and $49.4$~nm for the Au spacer. We note that the collector thickness obtained from the microscopy is somewhat smaller than the one used in the main manuscript for the data fitting. However, because the independent parameter utilized in the fitting routine is the ratio $D_{\rm Fe}/d^2$, this deviation can be mitigated by adjusting $D_{\rm Fe}$. It is known, for example, that thin Fe films of up to 24 monolayers demonstrate significantly lower stiffness (160 meV~$A^2$ as compared to 280 meV~$A^2$ for the bulk Fe) \cite{ProkopPRL09}. In our case, the collector thickness $d$ obtained using the TEM microscopy results in $D\approx~225$~meV~$A^2$, in between of these two values. Moreover, the lattice mismatch at the MgO/Fe interface leads to the appearance of dislocations and subsequent strain in the Fe film, which can also modify the effective magnon stiffness $D$. This strain could also induce the stiffness anisotropy resulting in the inequality of $D$ for the magnons with their $k$-vectors parallel (as in Ref.~\cite{ProkopPRL09}) and perpendicular (as in our work) to the film surface.

\section*{Supplementary Note 5: Realizing orthogonal magnetic configurations}

Static magnetic properties of the emitter and collector Fe layers were characterized by employing the magneto-optical Kerr effect (MOKE) measured independently. A thick (55 nm) Au spacer prevented magnetic coupling of the two Fe layers, as well as optical access to the collector from the emitter side and vice versa. As such, we performed simultaneous characterization of the magnetic states of the collector and emitter using MOKE with two laser beams. We note that thin Fe films have two in-plane orthogonal easy axes corresponding to the $\left\langle 100\right\rangle$ and $\left\langle 010\right\rangle$ directions (see also Ref.~\cite{ZhaoNJP09}). In our experiments, the sample was placed in such a way that one of these axes was in the incidence plane. We found that both magnetizations were switched to the opposite direction along one of the easy axes while the magnetic field in the longitudinal MOKE geometry $H_L$ was swept from -20 to 20 Oe. 

\begin{figure}[!h]
	\includegraphics[width=0.8\textwidth]{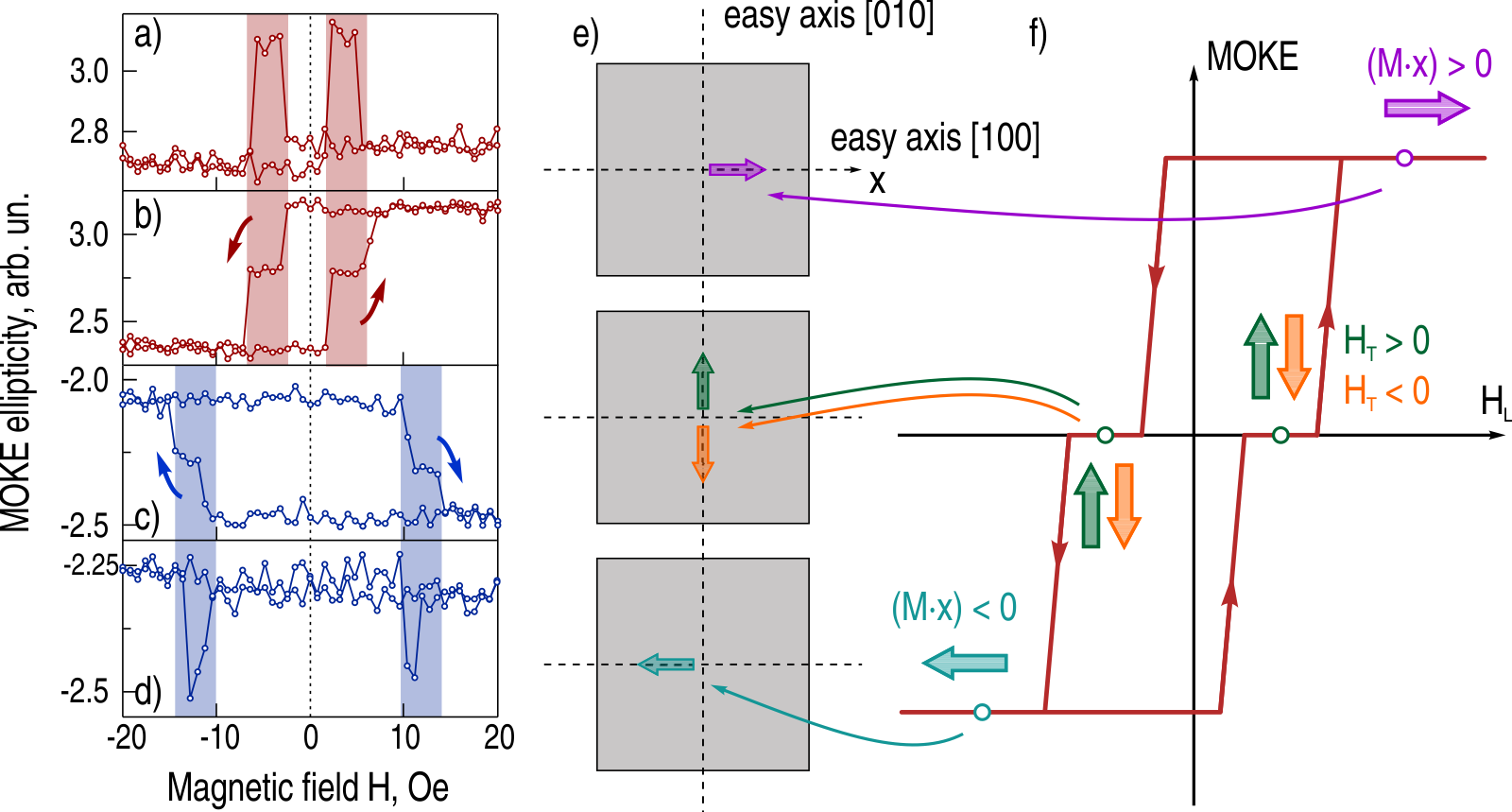}
	\caption{
	(a-b) Hysteresis loops measured with MOKE from the emitter (a) and the collector (b) Fe films.
	(c-d) Schematic of the magnetization switching using main $H_L$ and auxiliary $H_T$ magnetic fields applied perpendicular to each other along the two easy axes of magnetic anisotropy in Fe(100). The four coloured points at the sketch of the MOKE hysteresis loop (d) correspond to the four possible directions of magnetization (c).
	}
	\label{hystereses}
\end{figure}

In order to force the magnetization along the other easy axis we have applied a small auxiliary transversal magnetic field $H_T$ perpendicular to the main one. We found that for the values of $H_T$ as small as 4 Oe the switching of magnetization acquired a step in the middle (Fig.~\ref{hystereses},b-c) corresponding to the zero MOKE rotation and ellipticity. At the same time, when sweeping the transversal field while keeping the small $H_L$ on, we observed a sharp feature with a non-zero the MOKE rotation and ellipticity response (Fig.~\ref{hystereses},a,d). This indicates that the magnetization switching proceeds via an intermediate state which corresponds to the orthogonal orientation of the magnetization, along the second easy axis in the film. As such, when sweeping the main magnetic field $H_L$ across the hysteresis loop in the presence of the auxiliary field $H_T$ which could be positive or negative, we were able to attain any of the four orientations of the magnetization in the Fe film (Fig.~\ref{hystereses},e-f). 

It is seen in Fig.~\ref{hystereses},a-d that the hysteresis loops of the emitter and the collector have different widths, {\it i.e.} the coercive fields of the two films are unequal. This allows for the realization of a orthogonal configuration, where the magnetizations of the emitter and the collector are perpendicular to each other. For instance, if in the magnetic set-up used for the measurements shown in Fig.~\ref{hystereses},a-b longitudinal magnetic field of $\approx 12$ Oe was applied, the emitter and the collector were magnetized along the longitudinal and transversal axes, respectively. Varying the roles of the longitudinal and transversal magnetic fields (main or auxiliary) as well as their signs, it was possible to realize all 8 orthogonal magnetic configurations of the multilayer sample.

\end{document}